\documentclass[english, 12pt]{article}
\usepackage[T1]{fontenc}
\usepackage{babel}
\usepackage{ifsym}
\usepackage{geometry}
\usepackage{amsmath, amsthm, amssymb}
\usepackage{graphicx}
\usepackage{booktabs}
\usepackage{rotating}
\usepackage{setspace}
\usepackage{authblk}  
\usepackage{epigraph}
\usepackage{enumitem}   
\usepackage{url}
\usepackage{xcolor}
\usepackage[toc,page]{appendix}
\usepackage{threeparttable}
\usepackage[capposition=top]{floatrow}
\usepackage{setspace}
\usepackage{palatino}
\usepackage{pgfplots}
\usepackage{booktabs}
\pgfplotsset{compat=1.18}

\usepackage{bbm}
\usepackage{titlesec}
\usepackage[colorlinks]{hyperref}
\hypersetup{
	colorlinks=true,
	linkcolor=black,
	filecolor=blue, 
	anchorcolor=blue,
	citecolor=blue,
	urlcolor=black,
}

\usepackage[normalem]{ulem}
\usepackage{multirow}

\usepackage[style=authoryear-comp,natbib=true,maxnames=3, backend=bibtex]{biblatex}
\newtheorem*{theorem*}{Theorem}
\newtheorem{corollary}{Corollary}
\newtheorem{lemma}{Lemma}

\theoremstyle{definition}
\newtheorem{Remark}{Remark}

\newtheorem{Assumption}{Assumption}
\newtheorem{assumptionalt}{Assumption}[Assumption]
\theoremstyle{plain}            % <-- add this line
\newtheorem{Proposition}{Proposition}

\newenvironment{assumptionp}[1]{
  \renewcommand\theassumptionalt{#1}
  \assumptionalt
}{\endassumptionalt}
\theoremstyle{plain}            % <-- add this line
\newtheorem{propositionalt}{Proposition}[Proposition]

\newenvironment{propositionp}[1]{
  \renewcommand\thepropositionalt{#1}
  \propositionalt
}{\endpropositionalt}


\makeatletter
\usepackage{babel}

\usepackage{titlesec}
\titleformat*{\subsubsection}{\normalfont\fontsize{12}{17}\itshape}

\usepackage{footmisc}

\makeatother
\begin{document}
\title{Constraining to Motivate}
\pagenumbering{gobble}
\author{Liqun Liu\thanks{Corresponding author. School of International and Public Affairs, 
Shanghai Jiao Tong University, No.\ 1954 Huashan Road, Shanghai, China. 
Email: \text{liuliqunallen@gmail.com}.}}
%We thank Jia Chen, Songying Fang, Justin Fox, Giovanna Invernizzi, Barton Lee, Yu Mei, John Patty, Greg Sheen, for their incredibly helpful comments. All errors are mine. 

\date{}
\maketitle
\begin{abstract}
Effective decision-making often involves a principal delegating authority to better-informed agents. However, when agents' careers depend on how their decisions are perceived, they may avoid acquiring costly information and instead choose politically safe actions. We propose a simple solution: constrained delegation. By restricting the agent's choice set, the principal can reshape the reputational stakes attached to the remaining options and restore incentives to acquire information.  The mechanism applies broadly to settings where political or market-based incentives discourage information acquisition and shows how limiting discretion can, counterintuitively, improve decision quality.

\begin{doublespacing}
			
		\end{doublespacing}
			\vspace*{1cm}

            \noindent \textbf{Word Count:} 10731
			\end{abstract}

	\clearpage
                       
\noindent \textbf{Funding:} The author did not receive support from any organization for the submitted work.  

			\vspace*{1cm}

\noindent \textbf{Data Availability and Ethical Approval Statement:} This study did not involve the generation or analysis of datasets, nor did it require ethical approval, as the study is theoretical and mathematical in nature.

			\vspace*{1cm}

\noindent \textbf{Declaration:} The author declares no potential conflicts of interest with respect to the research, authorship, and/or publication of this article. 

During the preparation of this work the author used ChatGPT (OpenAI) and Claude (Anthropic) to assist with language polishing. After using this tool, the author reviewed and edited the content as needed and takes full responsibility for the content of the published article.

			\vspace*{1cm}
            \clearpage
\doublespacing
\pagenumbering{arabic}

\section{Introduction}
Modern governance and organizational design often
involve delegating authority to agents. Delegation helps overcome the limits of centralized control by enabling more informed and adaptive decision-making.\footnote{For reviews, see \citet{bendor2001theories,gibbons2013decisions}.} This idea underlies many practices, from public sector reforms to corporate management.

Yet delegated decision-making comes with hidden costs. Agents often face political or market-based incentives. While these incentives can align agent behavior with institutional goals, they may also backfire when agents prioritize career concerns over decision quality. In particular, if evaluations occur before outcomes fully materialize, agents may avoid acquiring costly information and instead settle on moderate, low-risk choices. For example, when given the option of either maintaining the status quo or experimenting with a spectrum of reforms, frequently rotated local officials often replicate centrally endorsed reforms rather than invest in tailored local studies.\footnote{\citet[pp.10--13]{heilmann2008policy} shows that local officials often replicate centrally endorsed models  to obtain experimental point status for preferential central support and better career prospects.} In essence, career concerns distort not only choices but also  the incentive to acquire information. Why does this motivation problem arise, and how can it be addressed?

\textbf{Our approach.} We study the motivation problem in a principal-agent framework that builds on \citet{maskin2004politician} and \citet{fox2007government}. Our novelty is to combine strategic delegation with costly information acquisition under career concerns. The principal first designs the set of permissible actions. The agent then decides whether to pay for information about the state and selects an action from the delegated set.  The agent can be either a \textit{congruent} or a  \textit{noncongruent} type, unknown to the principal. The congruent type shares the principal's preference, while the noncongruent type prefers an extreme option, such as maintaining the status quo in reform policymaking. Finally, after observing the agent's decision but before the outcome is realized, the principal updates beliefs about the agent's type and decides whether to reward or punish.

With full discretion, careerist agents often pander to moderate options  without acquiring information. Extreme options, typically associated with the noncongruent types, can hurt the agent's reputation particularly when outcomes are unobservable. Even if information reveals such choices to be optimal, they may still be seen as likely coming from the noncongruent type and invite dismissal.  This career risk undermines the incentive to acquire information (see also \citet{maskin2004politician}). Thus, agents settle on the moderate option to minimize ex ante policy risks without being blamed for acting conservatively.

We propose a  simple solution: \textit{constrained delegation}. By eliminating an extreme option, like removing the status quo in reform policymaking, the principal redefines which remaining options carry reputational risk. This change in reputational stakes can restore the congruent type's incentive to acquire information under career concerns.

Consider again a reform setting. Eliminating the status quo implies that change is expected. With only moderate and radical reforms available, the moderate option now appears overly conservative and invites replacement. An uninformed agent then faces two unappealing options: propose a radical reform, signaling progressiveness but inviting high policy risks; or propose a moderate reform, reducing policy risks but inviting replacement for acting conservatively. But there is a third way. By acquiring information, the agent can tailor the choice to actual conditions. This still entails career risk (rightly choosing the moderate policy may appear overly cautious ex ante), but it creates a more balanced tradeoff. 
Rather than being trapped between poor policy with career safety and better policy with career termination, an informed agent can escape this dilemma by designing strictly better policy while partially offsetting the career risks, albeit at an informational cost. Constrained delegation, by redefining what counts as extreme, makes information acquisition more attractive.

The motivation effect of constrained delegation extends beyond the reform
example. We formalize this intuition in a three-alternative environment and
characterize which action restrictions succeed in motivating information
acquisition. The results demonstrate that constrained delegation succeeds exactly when it destroys
the safety of the uninformed careerist's safe action.\footnote{The supplementary appendix extends this mechanism to richer action spaces. Informative equilibria are restored when action restriction reaches the ex ante optimal action, and the appendix characterizes when such a deep restriction remains profitable for the principal.}  Our mechanism differs from that of \citet{szalay2005economics}. In that framework, eliminating the compromising option motivates information acquisition by increasing the policy cost of remaining uninformed. In our setting, career concerns themselves create the motivation problem. Constrained delegation restores incentives by reshaping the career rewards and punishments attached to different actions rather than their direct policy payoffs.

Our model highlights career concerns as a source of demotivation when agents face political or market-based incentives. Relatedly, \citet{lee2025feigning} shows how careerist politicians may publicly support popular policies without exerting real effort to implement them. While \citet{lee2025feigning} models moral hazard 
in policy execution, we focus on moral hazard in acquiring information before decision-making. Constrained delegation offers a way to address this problem in agency relationships: instead of
relying on central control or full discretion, principals can restructure the choice set to alter agents' strategic calculus. One application is the use of special economic zones and other pilot programs. In China's ``experimental points'' \citep{heilmann2008policy}, local officials were authorized to implement reforms without the option of maintaining the status quo. Success in these zones often translated into career advancement, aligning political incentives with informed experimentation. While geography, investment, and central support mattered, the institutional design itself encouraged more active and informed policymaking.

Our principal-agent model sits between models of pandering \citep{maskin2004politician,fox2007government}, delegation \citep{holmstrom1980theory, melumad1991communication, szalay2005economics}, and  accountability \citep{fearon1999electoral, patty2021ex}.
It combines ex ante control (strategic delegation) and ex post oversight (accountability) to study how to select and control political agents without monetary transfers. Prior work highlights the costs of delegation or the policy distortions caused by pandering; we instead show how ex ante control can raise the value of information and restore incentives to acquire it. More broadly, our analysis illustrates how institutional design can motivate careerists when explicit contracts are infeasible.

\section{Model}
A principal $P$ (she) delegates policymaking authority to an agent $A$ (he). The state of the world $\omega \in \Omega := \{0,1,r\}$ is drawn with probabilities $p_\omega \in (0,1)$, where $r > 2$ and $\sum_\omega p_\omega = 1$. Before choosing a policy $x \in D \subseteq \Omega$, where $D$ is the set of policies permitted by $P$, $A$ may acquire information about the true state $\omega$. The correct policy for  the principal's interest is the one that matches the state ($x=\omega$).  For clarity of exposition, we use reform policymaking as a running interpretation: $0$ is interpreted as the ``status quo,'' $1$ as a ``moderate reform,'' and $r$ as a ``radical reform.''\footnote{All three labels are purely interpretive: formally, $x=0$ is the
noncongruent type's bliss action, $x=1$ the ex ante optimal action, and
$x=r$ the action valuable in unusual states. None of our results depend on
these interpretations. We adopt the reform labels because our leading
applications concern reform environments.}  $r > 2$ captures the possibility that an extreme response might be needed to tackle unusual situations. This structure extends that of \citet{fox2007government} by allowing for information acquisition within a richer action space. 

The agent has a private type, either \textit{congruent} ($t = c$), who shares the principal's preference for matching policy to state, or \textit{noncongruent} ($t = n$), who prefers the lowest action regardless of the state.\footnote{We follow \citet{fox2007government} here. Substantively, congruent types may be policy entrepreneurs or reformists aiming to establish a policy legacy, while noncongruent types could be political conservatives influenced by interest groups or ideologues resistant to change. In Section \ref{Noncongruence} of the Appendix, we show that our results remain robust when noncongruent types have state-dependent or ``radical'' preferences.} The prior is $P(t=c)=\pi\in (0,1)$.  The principal and the congruent type both receive $v_c = -(x - \omega)^2$; the noncongruent type receives $v_n = -x^2$. Both types also receive office rent $R > 0$ upon retention. We will denote a congruent agent who has observed state $\omega$ as type $(c, \omega)$, and a noncongruent agent simply as type $n$ (the noncongruent type's payoffs are state-independent, so his information state is irrelevant for behavior).

The game has three periods. In the first period, $P$ chooses the delegation set $D\subseteq\Omega$. After that, Nature draws the state $\omega$ and the agent's type $t$ from their prior distributions. In the second period, $A$ observes $t$, and decides whether to exert effort at a cost $k>0$ to observe $\omega$. Later, $A$ chooses a policy $x\in D$. In the third period, $P$ observes only the policy $x$ without observing the state $\omega$ or whether the agent acquired information.\footnote{If the principal observes the information acquisition decision, then this decision rather than policy choice signals the agent's type. This setup would undermine the purpose of this article. } $P$ must decide whether to retain or replace the agent based on $x$.  Conditional on replacement, $P$ will  randomly select another agent from the same pool of candidates. $P$ gets a utility of $0$ from selecting a noncongruent type and $B>0$ from selecting a congruent type.

We characterize perfect Bayesian equilibria (PBE) where the principal uses pure retention strategies. This allows us to present the model's insights in the simplest form. Remark \ref{Re~cost} in the appendix shows in some parameter regions only mixed strategy equilibria exist, but they share similar equilibrium properties and strategic implications. The Online Supplementary Appendix characterizes equilibria under mixed retention strategies and shows that the informative-equilibrium region comparison of the main paper is preserved.\footnote{The Online Supplementary Appendix also extends the analysis to richer action spaces (with $x = 1$ ex ante optimal, and with the ex ante optimum interior), where the motivation result of Corollary~\ref{cor:sq vs full} extends and the welfare comparison holds under a condition characterized there.} Moreover, since numerous off-path beliefs are possible, we focus on equilibria that survive the divinity criterion. Intuitively, this implies that a deviation is most likely to come from the type that benefits most from it.

Our analysis proceeds in two steps: (1) fix $D$ and characterize the equilibrium behavior under that delegation set, and (2) identify conditions under which eliminating the status quo maximizes the principal's expected payoff. Here we focus on intuition, and leave technical details to the appendix.

Before proceeding, we impose several assumptions. 
\begin{Assumption}\label{Assumption: Ideal}
    $p_0=p_r=p\in (0,\frac{1}{2})$. 
\end{Assumption}
Assumption \ref{Assumption: Ideal} ensures that when uninformed, the principal's preferred policy is $x=1$. This reflects the appeal of policy compromise in minimizing ex ante risk. In Section \ref{Asymmetry} of the Appendix, we show that this symmetric distribution assumption does not alter players' strategic behaviors and, therefore, does not affect the core findings of our analysis.
\begin{Assumption}\label{Assumption: Office}
    $R \in (1, (r-1)^2)$.
\end{Assumption}
Assumption \ref{Assumption: Office} implies that  the office rent $R$ is moderate. It prevents agents with ideal policy $x^*\neq r$  from pandering to $x=r$ for retention.   A very small $R$ does not affect the agent's policymaking behavior; a very large $R$ or $r< 2$ encourages pandering to the ``congruent action'' to secure retention without acquiring information. 

\begin{Assumption}\label{Assumption: Cost}
    $k > k^0 := p(r-1)^2$.
\end{Assumption}
Assumption \ref{Assumption: Cost} restricts attention to the parameter region in which full delegation fails to motivate information acquisition. This is the region in which the motivation problem is substantive: for $k \leq k^0$, the congruent type acquires information even under full delegation, and there is nothing to fix. We maintain this assumption throughout the main analysis.

\section{Analysis}
The principal's first best is that the agent acquires and acts on policy-relevant information. But the noncongruent type never does so: his ideal action is always the lowest, so information cannot improve his policy payoff. Politically,
the principal does not observe information acquisition, so acquiring it cannot affect retention. Whatever an informed noncongruent type would do, he can replicate while uninformed to save the cost $k$.

 \begin{lemma}\label{L1}
In any equilibrium, the noncongruent type does not acquire information. 
\end{lemma}
\begin{proof}
    All proofs are in the Appendix.
\end{proof}

Thus, the principal can only hope that the congruent type acquires information and chooses the correct policy. For any subgame induced by a delegation set, we call an equilibrium \textit{informative} if the congruent type acquires information along the equilibrium path and \textit{noninformative} otherwise.  Lemma \ref{L2} outlines the key features of informative equilibria and highlights the principal's  tradeoff: informed policymaking comes at the cost of losing control over misaligned types. 
 
\begin{lemma}\label{L2}
Generically, in any informative equilibrium, the noncongruent type will always be replaced, and thus  chooses the lowest action within the delegation set. 
\end{lemma}
 
The logic is straightforward. If the noncongruent type were retained, he would choose the lowest action that secures retention. However, the congruent type utilizes information and sometimes chooses other actions. As a result, the lowest action becomes more associated with noncongruence, lowering the principal's posterior and leading her to replace rather than retain. Lemma \ref{L2} allows us to characterize equilibrium  behavior under each delegation strategy. 

%Importantly, informative equilibria under pure retention are unique. By Lemma \ref{L2}, the noncongruent type chooses the lowest available action and is replaced, while every other action is associated only with the congruent type and therefore retained.\footnote{For higher actions that are off path, the same conclusion follows from the divinity criterion. The only type who benefits from deviating is the congruent type observing the corresponding state, so the posterior equals one and retention follows.} Thus, the retention rule is pinned down by sequential rationality together with the divinity refinement, without invoking the tie breaking convention. The only equilibrium multiplicity in the model is therefore the coexistence of informative and uninformative pooling equilibria, which we characterize under each delegation strategy below.

\subsection{Baseline: full delegation}\label{Section:baseline}
Consider full delegation, i.e., $D = \{0,1,r\}$. The principal's policy payoff is maximized if the congruent type acquires information and chooses $x=\omega$. This requires that the value of information exceeds its cost.
 
However, career concerns weaken this incentive. Choosing $x = 0$, even if correct ex post, signals noncongruence and leads to replacement. To avoid  this, the agent may prefer $x = 1$, which appears moderate and safe.  Since the uninformed congruent type always chooses $x = 1$, information has value only when $\omega = r$. To induce information acquisition requires $k\leq k^0 := p(r-1)^2$, where $k^0$ is the value of information with career concerns. By contrast, an agent unconcerned with retention (e.g., $R = 0$) values information at $k^U := p((r - 1)^2 + 1)$ (see also \citet{maskin2004politician}). Under Assumption \ref{Assumption: Cost}, $k > k^0$ and full delegation fails to motivate information acquisition.

\begin{Proposition}\label{Prop~UD}  Suppose Assumption~\ref{Assumption: Cost} holds. An informative equilibrium does not exist, and a unique equilibrium exists, in which (1) both types of the agent choose $x=1$ without acquiring information and (2) the principal retains after observing $x\geq 1$.
\end{Proposition}
 
In equilibrium, both types pander to the ex ante popular policy of ``moderate reform'' ($x=1$).  For the noncongruent type, $x = 1$ secures retention at tolerable policy cost. For the congruent type, no profitable deviation exists: uninformed deviations lower his policy payoff without career gains, and information is worthwhile  only when $k \leq k^0$.

Informative equilibria are impossible. They require the noncongruent type to choose $x=0$ and be replaced (Lemma \ref{L2}). But to deter him from deviating to $x = 1$, the principal must commit to replacing anyone who chooses it. This is not credible: the congruent type, if informed that $\omega = 1$, can signal congruence by choosing $x=1$.\footnote{If the moderate option is on path, the principal's posterior is $P(t=c|x=1)=1$; if it is off path, the posterior is still $P(t=c|x=1)=1$ due to the divinity condition, as the congruent type observing $x=1$ is most likely to benefit from the deviation.} Thus, the principal cannot credibly penalize $x = 1$, and informative equilibria collapse.

The principal commits to $D$ ex ante but cannot commit to a retention rule, since retention must be sequentially rational. This is why she cannot simply punish $x=1$ under full delegation. She thus obtains the same outcome she would have chosen herself. Worse, for $k \in (k^0, k^U)$, an unaccountable agent would have acquired and used information, but a careerist does not.

\subsection{Constrained delegation}\label{Section:constrain}
Throughout this section, the delegation set should be understood as the agent's realistically available actions rather than a literal menu. Institutional arrangements need not formally prohibit an action; it is enough that they make it sufficiently unattractive that the agent never chooses it in equilibrium. Financial penalties, public commitments, and legal or institutional constraints can all generate such restrictions. We now examine the three possible restrictions of the action set. Assumption \ref{Assumption: Cost} is assumed throughout.

\subsubsection{Eliminating the status quo}\label{Section:sq}
Eliminating the status quo ($x = 0$) corresponds to institutional arrangements that make inaction politically or institutionally untenable. It encompasses reform mandates, pilot programs,  or public commitments by senior leaders to change course. At first glance, such a restriction may appear inconsequential. Since neither type chooses the status quo under full delegation, removing it should have no effect on equilibrium behavior.

But this view overlooks a key effect: eliminating the status quo allows the principal to \textit{credibly} replace agents choosing the moderate option. The rationale mirrors practices in special economic zones, where unambitious reform decisions (like replicating established models) may be perceived as conservative and expose agents to dismissal. This reshapes the reputational stakes across options and encourages information acquisition. 
 
Suppose the principal retains only if $x = r$. The noncongruent type, who finds $x = r$ strictly dominated ($R < r^2 - 1$), will choose $x = 1$ and give up retention. An informed congruent type will choose $x = r$ when $\omega = r$ and $x = 1$ otherwise. The principal's retention rule is sequentially rational: $x = r$ is uniquely associated with (informed) congruent types, while $x = 1$ is chosen more frequently by the noncongruent type, driving the posterior there below $\pi$. 
The congruent type finds information acquisition worthwhile whenever its cost falls below the adjusted value of information, denoted $k^S= \min\{\, p(R + (r-1)^2),\;
p[(r^2 - 1) - R] + (1-2p)[(r-1)^2 - R] \,\}$.
 
\begin{Proposition}\label{prop:sq} 
When the principal eliminates the status quo, i.e., $D = \{1, r\}$, an
informative equilibrium exists if and only if $k \leq k^S$. In this
equilibrium, the noncongruent type chooses $x = 1$ without acquiring
information and is replaced; the congruent type acquires information,
chooses $x = r$ if and only if $\omega = r$ and $x = 1$ otherwise, and
is retained if and only if $x = r$.

A pooling equilibrium in which both types choose $x = 1$ and are
retained exists if and only if $k > k^0$; hence for
$k \in (k^0, k^S]$ it coexists with the informative equilibrium. For
$k > k^S$, one further uninformed equilibrium arises: separation (the
noncongruent type at $x = 1$, replaced; the congruent type at $x = r$,
retained) if $R \geq \underline{R} := p(r^2 - 1) + (1-3p)(r-1)^2$, and
pooling at $x = 1$ with replacement if $R \leq \underline{R}$.
 \end{Proposition}

Eliminating $x = 0$ thus changes equilibrium behavior even though neither type chooses it under full delegation. Once $x = 1$ becomes the lowest available action, it is chosen relatively more often by the noncongruent type and can therefore be credibly punished, while $x = r$ reveals congruence. The restriction thus makes the uninformed careerist's safe harbor punishable and changes the congruent type's value of information from $k^0$ to $k^S$.

\begin{lemma}\label{lemma: cutoff}
    $k^S>k^0$ for a nondegenerate set of parameters $(p,r,R)$.
\end{lemma}
For $k\in(k^0,k^S]$, eliminating the status quo therefore
supports an informative equilibrium while full delegation does not, although the pooling
equilibrium continues to coexist.  Corollary~\ref{cor:sq vs full} shows the principal is weakly better off regardless of selection.

\subsubsection{Eliminating the moderate option}\label{Section:m}
We now consider removing the moderate option ($D = \{0, r\}$). Substantively, this restriction captures institutions that eliminate compromise, forcing a choice between preserving the status quo and pursuing a more ambitious alternative.  We characterize the equilibrium and then compare it to a closely related setting in \citet{szalay2005economics}.

\begin{Proposition}\label{prop:m}
When the principal eliminates the moderate option, i.e., $D=\{0, r\}$, there exists $k^M>0$ such that an informative equilibrium exists if and only if $k\leq k^M$. In this equilibrium, the noncongruent type chooses $x=0$ without acquiring information and is replaced; the congruent type acquires information, chooses $x=\omega$ for $\omega\in \{0,r\}$, and at $\omega=1$ chooses $x=r$ if $R \geq r(r-2)$ and $x=0$ otherwise; the principal retains if and only if $x=r$.

A pooling equilibrium in which both types choose $x=0$ and are retained exists if and only if $k > p r^2$; it coexists with the informative equilibrium on a nonempty interval if and only if $R < \frac{(1-2p)r(r-2)}{1-p}$. The remaining uninformative equilibria arise only for $k > k^M$ and are characterized in the appendix.

\end{Proposition}

This restriction also induces information acquisition, but through a different margin from Section~\ref{Section:sq}. With the moderate option removed, the only action that secures retention is $x=r$. The noncongruent type finds $x=r$ strictly dominated by $x=0$ followed by replacement, and therefore gives up retention. The congruent type, by contrast, can secure retention only by choosing $x=r$, which makes information valuable whenever its cost falls below $k^M$.

This comparison clarifies the connection to \citet{szalay2005economics}. In Szalay, eliminating the compromise option raises the direct policy cost of remaining uninformed: a juror restricted to ``guilty'' or ``not guilty'' has stronger incentives to weigh evidence than one who can choose an intermediate no-decision option. Here, by contrast, the restriction works through career concerns. It changes the retention consequences attached to the remaining actions, making $x=0$ a signal of noncongruence and $x=r$ a signal of congruence.

The two restrictions in our framework differ in the \textit{strength} of the incentives they generate. For clarity of exhibition, we focus on the following scenario:
\begin{lemma}\label{lemma: SvM}
Suppose $R \geq r(r-2)$. Then $k^S > k^M$ if and only if $2R > 2r-1$ and $p < (1-2p)[(r-1)^2 - R]$.
\end{lemma}

Lemma~\ref{lemma: SvM} shows that neither restriction uniformly dominates the other. For a nondegenerate range of parameters, $k^S>k^M$, so eliminating the status quo motivates information acquisition over a wider range of information costs than eliminating the moderate option; outside that range, the ranking reverses. The comparison therefore depends on the office rent $R$, the policy stakes generated by radical option, and the likelihood of the relevant states. Figure~\ref{fig:thresholds} plots $k^S$ and $k^M$ against the full delegation benchmark $k^0$.

\begin{figure}[t]
\centering
\caption{Comparing Cutoffs}
\begin{tikzpicture}
\begin{axis}[
    width=12cm, height=8cm,
    xlabel={Office rent $R$},
    ylabel={Information cost cutoff},
    xmin=3, xmax=4,
    ymin=0.6, ymax=1.6,
    xtick={3, 3.667, 4},
    xticklabels={$r(r-2)$, $R^{*}$, $(r-1)^2$},
    ytick={0.8, 1.067},
    yticklabels={$k^{0}$, $k^{S}=k^{M}$},
    axis lines=left,
    legend style={draw=none, fill=none, at={(0.98,0.98)}, anchor=north east, font=\small},
    samples=200,
    domain=3:4,
]
% k^0 = p(r-1)^2 = 0.8 (full-delegation cutoff)
\addplot[gray, dashed, thick] {0.8};
\addlegendentry{$k^{0}$ (full delegation)}

% k^M = p(r^2 - R)
\addplot[red!70!black, thick] {1.8 - 0.2*x};
\addlegendentry{$k^{M}$ (eliminate moderate)}

% k^S = min(A, B)
\addplot[blue!70!black, thick] {min(0.2*x + 0.8, 4 - 0.8*x)};
\addlegendentry{$k^{S}$ (eliminate status quo)}

\addplot[gray, dashed] coordinates {(3.667, 0.6) (3.667, 1.067)};
\addplot[mark=*, mark size=2pt, only marks, black] coordinates {(3.667, 1.067)};
\end{axis}
\end{tikzpicture}

\label{fig:thresholds}

\floatfoot{\textit{Notes}: This figure compares information acquisition cutoffs at $r=3$, $p=0.2$. In this parameter region, both constrained strategies raise the cutoff above the full-delegation baseline $k^{0}$. Eliminating the status quo ($k^S$) dominates eliminating the moderate option ($k^M$) when $R<R^*$; the comparison reverses for $R>R^*$.}
\end{figure}

\subsubsection{Eliminating the radical option}\label{Section:radical}
The remaining restriction is $D = \{0, 1\}$, which removes the radical option. Substantively, this restriction represents institutions that rule out the most ambitious actions.
Unlike eliminating the status quo, this restriction does not reshape reputational stakes in a useful way: the moderate option $x = 1$ remains the safe harbor, and the noncongruent type can still pander to it. Career concerns therefore continue to suppress information acquisition.
 
\begin{Proposition}\label{prop:ext}
When the principal eliminates the radical reform, i.e., $D = \{0, 1\}$, no informative equilibrium exists for any $k>0$. The unique equilibrium pools both types at $x = 1$ without information acquisition; the principal retains after $x=1$ and replaces after $x=0$.
\end{Proposition}
The pooling outcome inherits the logic of full delegation: $x=1$ remains a safe choice for both types, and the principal retains because she cannot tell them apart. What changes is that the informed congruent type can no longer use information at $\omega=r$, since $x=r$ is unavailable. The restriction removes the action for which information is most valuable while leaving the safe harbor intact. It therefore binds on the wrong margin. Eliminating the radical option cannot address the motivation problem. We treat it briefly here and defer the formal argument to the appendix.

\subsection{Choosing among delegation strategies}\label{Section:endogenous delegation}
\subsubsection{Equilibrium patterns}
Having characterized how each restriction shapes behavior, we now ask which the principal should choose. Table \ref{tab:delegation-summary} collects the equilibrium outcomes across the four delegation sets.\footnote{Singletons are trivially suboptimal: forced to fix one policy, the principal would choose her ex ante optimum $x=1$, which merely reproduces the pooling outcome, so we consider only $|D|\geq 2$.}

\begin{table}[ht]
\centering
\begin{threeparttable}
\begin{tabular}{lllll}
\toprule
Delegation set $D$ & Informative iff & Noncongruent plays & Congruent plays & Sorting \\
\midrule
$\{0,1,r\}$ (full)
  & never\tnote{a}
  & $x=1$, retained
  & $x=1$, uninformed
  & none \\
$\{1,r\}$ (no status quo)
  & $k \leq k^S$
  & $x=1$, replaced
  & $x=\omega$ at $\omega=r$;\; $x=1$ o.w.
  & at $x=r$ \\
$\{0,r\}$ (no moderate)
  & $k \leq k^M$
  & $x=0$, replaced
  & $x=\omega$ at $\omega\in\{0,r\}$\tnote{b}
  & at $x=r$ \\
$\{0,1\}$ (no radical)
  & never\tnote{c}
  & $x=1$, retained
  & $x=1$, uninformed
  & none \\
\bottomrule
\end{tabular}
\caption{Equilibrium outcomes under the four delegation sets with $k > k^0$}
\label{tab:delegation-summary}
\begin{tablenotes}\footnotesize
\item[a] By Assumption~3 ($k > k^0$); an informative equilibrium
requires $k \leq k^0$ (Proposition~1). Lemma~\ref{lemma: cutoff} shows $k^S > k^0$
for a nondegenerate parameter set, so the region $k \in (k^0, k^S]$
is where constrained delegation succeeds and full delegation fails.
\item[b] At $\omega = 1$: $x = r$ if $R \geq r(r-2)$, $x = 0$
otherwise (Proposition~3). Lemma~\ref{lemma: SvM} characterizes when
$k^S \gtrless k^M$.
\item[c] For any $k > 0$: under the unique sequentially rational
retention rule, informed and uninformed play coincide, so
information has zero value (Proposition~4).
\end{tablenotes}
\end{threeparttable}
\end{table}
The pattern is clear: a restriction motivates information acquisition exactly when it destroys the safety of the uninformed careerist's safe harbor. This can happen by removing that action outright, as under $D=\{0,r\}$, or by leaving it in place while making it credibly punishable, as under $D=\{1,r\}$. By contrast, $D=\{0,1\}$ binds on the wrong margin: it removes the action most valuable to an informed agent while leaving the safe harbor intact.

\subsubsection{Constrained delegation versus full delegation}
We now compare their welfare implications. Three comparisons are of particular interest: constrained versus full delegation, alternative forms of constrained delegation, and constrained delegation versus delegating to an unaccountable agent. Throughout, the welfare comparisons are shaped by the prior probability of congruence $\pi$, the likelihood that the status quo is optimal $p_0$, the urgency of radical reform $r$, and the office rent $R$, which determines which incentive constraint binds.

The first comparison establishes that for suitable parameter restrictions, eliminating the status quo never makes the principal worse off than full delegation when the latter suffers from an information problem, and strictly improves her payoff whenever it succeeds in motivating information acquisition.
 
\begin{corollary}\label{cor:sq vs full}
 Suppose $k \in (k^0, k^S]$. Relative to full delegation, eliminating the status quo yields the same payoff for the principal when the pooling equilibrium is selected, and a strictly higher payoff when the informative equilibrium is selected.
\end{corollary}
The restriction never leaves the principal worse off than full delegation.\footnote{The weak-dominance result relies on the baseline action set $D=\{0, 1,r\}$. It leaves no sufficiently attractive action above $x=1$ for the congruent type to pander to at low states. The informative equilibrium therefore preserves the full-delegation policy at $\omega\in\{0,1\}$ and improves it at $\omega=r$. With richer action spaces, nearby higher actions may induce upward pandering, so welfare dominance is conditional. See Online Appendix Part I.} If the pooling equilibrium is selected, both types pool at $x=1$ and the principal obtains the same outcome as under full delegation. If the informative equilibrium is selected, she gains on both policy and screening: the congruent type matches the state at $\omega=r$, and choosing $x=r$ reveals congruence, allowing her to retain an agent who would otherwise be pooled with the noncongruent type. Eliminating the status quo therefore preserves the full delegation outcome when information acquisition fails, while creating the possibility of a strictly better outcome when it succeeds.

\subsubsection{Choosing among constrained delegation strategies}
We next compare alternative forms of constrained delegation. Among the three, eliminating the radical option is dominated: by Proposition \ref{prop:ext}, it cannot motivate information acquisition. The substantive comparison is therefore between eliminating the status quo and eliminating the moderate option. When the status quo is unlikely to be optimal, excluding it entails smaller efficiency loss than excluding the moderate option.

To compare the two nontrivial restrictions on welfare grounds, we now allow a general prior $(p_0,p_1,p_r)$. This makes the policy cost of excluding the status quo explicit. Section~\ref{Asymmetry} verifies that equilibrium behavior under each delegation set is unchanged up to this relabeling; the information thresholds are understood as their
general-prior counterparts characterized there.

We compare the informative equilibrium under each restriction.
\begin{corollary}\label{cor:sq vs m}
There exists $\bar{B} > 0$ such that for all $B \leq \bar{B}$,
eliminating the status quo yields higher welfare than eliminating
the moderate option if and only if the status quo is unlikely:
$p_0 \leq \bar{p}_0^{\mathrm{m}}$, where
\begin{equation*}
\bar p_0^{\,m}(p_1,p_r,r,\pi;B) =
\begin{cases}
p_1+(1-\pi)p_r(2r-1), & R<r(r-2),\\[4pt]
\pi p_1(r-1)^2+(1-\pi)\big(p_1+p_r(2r-1)\big)-\pi(1-\pi)p_1B, & R\ge r(r-2).
\end{cases}
\end{equation*}
\end{corollary}

The threshold $\bar{p}_0^{\mathrm{m}}$ balances the policy cost of forbidding the status quo against the gain from disciplining the noncongruent type. When $R < r(r-2)$, the two restrictions induce identical selection payoffs. When $R \geq r(r-2)$, eliminating the moderate option carries an additional selection advantage: it also identifies congruence at $\omega=1$. Eliminating the status quo is therefore favored when the status quo is unlikely to be optimal (low $p_0$) and when radical reform is urgent (high $r$, which widens the wedge between informed and uninformed play at $\omega=r$); when the status quo is likely optimal, the policy cost dominates and eliminating the moderate option is preferred.

 \subsubsection{Constrained delegation versus unaccountable agents}
A different response to the motivation problem is to bypass career concerns altogether by delegating to an unaccountable agent ---a technocrat, a lame duck, or any official whose retention does not depend on the principal's decision. Such an agent acquires information whenever its value exceeds its cost, but the principal loses her ability to screen out misaligned types. This is the response advocated by \citet{maskin2004politician}. We compare it to our mechanism. 
 
We assume $k \in (k^0, \min\{k^S, k^U\})$: full delegation fails, but both constrained delegation and delegation to unaccountable agents motivate information acquisition. If $k > k^U$, delegating to unaccountable agents is pointless as they will not acquire information.\footnote{Under the general prior,
$k^U = p_0 + p_r(r-1)^2$: an unaccountable agent when uninformed plays
the ex ante optimal $x = 1$, so information gains $1$ at $\omega = 0$ and $(r-1)^2$ at
$\omega = r$.} In addition, when the principal delegates to an unaccountable agent, his tenure is not conditioned on the observed action. The principal therefore obtains no screening value from the delegation decision and receives $\pi B$ from a random replacement.
 
\begin{Proposition}\label{prop:nonaccountability}
Suppose $k \in (k^0, \min \{k^S, k^U\})$. Eliminating the status quo yields higher welfare than delegating to unaccountable agents if and only if $p_0\leq (1-\pi)(p_r(2r-1)+p_1)+\pi (1-\pi)p_rB$.
\end{Proposition}
Relative to an unaccountable agent, eliminating the status quo incurs a policy loss at $\omega=0$, but disciplines the noncongruent type and preserves the principal's ability to screen agents. It is therefore favored when the status quo is unlikely to be optimal, when preference conflict makes discipline valuable, or when the value of screening is high. Delegation to an unaccountable agent is more attractive when preserving policy flexibility at $\omega=0$ is sufficiently important. Whereas \citet{maskin2004politician} restore information acquisition by removing career incentives, constrained delegation restructures those incentives while preserving the disciplining and screening roles of accountability.

\section{Conclusion}
A crucial challenge to decision quality is the lack of policy relevant information, and delegation is often used to address it. But career concerned agents tend to choose politically safe actions rather than acquire the information needed to make good decisions. Career concerns therefore distort not only which policies agents choose, but also whether they acquire information before choosing.

We propose a simple solution to this motivation problem. By restricting the agent's choice set, the principal reshapes the reputational stakes attached to the remaining actions and can restore the incentive to acquire information. This logic helps explain institutional arrangements that remove the option of preserving the status quo and require officials to choose among more consequential alternatives.

More broadly, the paper offers a different perspective on constrained delegation. Restrictions on discretion are usually understood as a means of limiting agency loss. Our analysis identifies another role: constrained delegation can improve decision quality by changing the incentives agents face before decisions are made. It therefore serves not only to control behavior, but also to motivate information acquisition.

 	\renewbibmacro{in:}{}
	\printbibliography
 \clearpage

\setcounter{page}{1}

\begin{appendices}
 
\section{Solution concept and refinement}
Let $D$ be a fixed delegation set with typical element $d$. An agent's information acquisition strategy, $\tau_t$, is a function of his type $t \in \{c, n\}$, where $\tau_t = 1$ indicates information acquisition and $\tau_t = 0$ indicates no information acquisition.
 
An agent's policy strategy is a single mapping
\[
\sigma_t : \Omega \cup \{\emptyset\} \;\to\; \Delta(D),
\]
where $\sigma_t(\omega)$ is the distribution over actions in $D$ chosen by an informed type-$t$ agent observing state $\omega$, and $\sigma_t(\emptyset)$ is the distribution over actions chosen by an uninformed type-$t$ agent. We write $\sigma_t(\omega)(d)$ for the probability that action $d$ is chosen.
 
The principal does not observe whether information was acquired or what state was realized. Conditional on observing action $d$, the probability that this action was taken by a type-$t$ agent is
\[
A_d(t) \;:=\; (1-\tau_t) \sigma_t(\emptyset)(d) \,+\, \tau_t \sum_{\omega \in \Omega} p_\omega\, \sigma_t(\omega)(d).
\]
The principal's posterior belief that the agent is congruent upon observing $d$ is
\begin{align*}
\mu_d:= P(t = c | x = d) = \frac{\pi A_d(c)}{\pi A_d(c) + (1 - \pi) A_d(n)}.
\end{align*}
The principal's strategy $\rho : D \to [0,1]$ maps each observed action to a probability of retention: $\rho(d)$ is the probability the principal retains the agent after observing action $d$. Pure retention corresponds to $\rho(d) \in \{0, 1\}$. An action $d$ is good (bad) news for retention if $\mu_d\geq \pi$  ($\mu_d< \pi$).
 
A perfect Bayesian equilibrium is characterized by measurable $\tau_t$, $\sigma_t$, and $\rho$ such that:
\begin{enumerate}
\item For each $t$ and each $d$ in the support of $\sigma_t(\emptyset)$ (when $\tau_t = 0$):
    \begin{align*}
        d\in \arg \max_{x\in D} \sum_\omega p_\omega v_t(x,\omega)+\rho(x)R.
    \end{align*}
For each $t$ and each $d$ in the support of $\sigma_t(\omega)$ (when $\tau_t = 1$):
    \begin{align*}
        d\in \arg \max_{x\in D} v_t(x,\omega)+\rho(x)R.
    \end{align*}
\item $\rho(d)=1 $ for $\mu_d>\pi$, $\rho(d)=0 $ for $\mu_d<\pi$, and $\rho(d) $ takes any value in $[0,1]$ for $\mu_d=\pi$.
\item Let  $V_t(0):=\max_{x\in D} \sum_\omega p_\omega v_t(x,\omega)+\rho(x)R$ and $V_t(1):=\sum_\omega p_\omega\max_{x\in D}  v_t(x,\omega)+\rho(x)R$ be a type-$t$ agent's expected continuation payoff after the information acquisition decision. Then   $\tau_t=1\Leftrightarrow k\leq V_t(1)-V_t(0)$.
\item The belief $\mu_d$ is determined by Bayes' rule whenever possible.
\end{enumerate}
 
The divinity refinement, proposed by \citet{banks1987equilibrium} and used in the treatment by \citet{fudenberg1991game}, places additional restrictions on the principal's off-path beliefs. Suppose an action $d'$ is off path. Let $\bar{\rho}(d',T,t):=E[\rho(d') | T,t]$ be the principal's mixed strategy best reply to $d'$ (retention probability) under beliefs concentrated on $T$ that make type $t$ strictly prefer $d'$ to his equilibrium strategy. Here $T\subset \{c,n\}\times (\Omega \cup \{\emptyset\})$, with the second coordinate indicating whether the agent has acquired information. If $\bar{\rho}(d',T,c)>\bar{\rho}(d',T,n)$, then the noncongruent type is more likely to deviate to $d'$, since the congruent type requires more retention compensation to benefit. We use this refinement in equilibrium construction.
 
\section{Technical results and omitted proofs}
Throughout this section, we mean ``the posterior'' by ``the posterior belief that the agent is congruent''.

\begin{proof}[Proof of Lemma \ref{L1}]
Fix a delegation set $D$. Suppose, toward a contradiction, that in some equilibrium the noncongruent type acquires information: $\tau_n = 1$. Let $\sigma_n(\omega)$ denote his informed strategy. Consider the alternative strategy $(\tau_n', \sigma_n')$ in which the noncongruent type does not acquire information ($\tau_n' = 0$) and plays the following strategy
\[
\sigma_n'(\emptyset)(d) \;:=\; \sum_{\omega \in \Omega} p_\omega\, \sigma_n(\omega)(d), \qquad d \in D.
\]
This deviation generates the same distribution over actions as $(\tau_n, \sigma_n)$, so by the definition of $A_d(t)$ it leaves $A_d(n)$ unchanged for every $d$. The principal's posterior $\mu_d$ and retention decision $\rho(d)$ are therefore unchanged. Since the noncongruent type's policy payoff $v_n(x,\omega) = -x^2$ is state-independent, his expected policy payoff is also unchanged. The deviation thus yields the same continuation payoff while saving the information cost $k > 0$, contradicting equilibrium.
\end{proof}

\begin{proof}[Proof of Lemma \ref{L2}]
Suppose the congruent type has acquired information at positive cost $k > 0$. He must play an informative strategy on the equilibrium path to be sequentially rational. Let $W$ be the set of actions that the informed congruent type plays with nondegenerate probabilities; informativeness implies $|W| \geq 2$.
 
Let $D_R \subseteq D$ denote the set of actions after which the principal retains, i.e., $D_R := \{d \in D : \rho(d) = 1\}$. We argue that the noncongruent type chooses an action outside $D_R$, i.e., is replaced.
 
Suppose for contradiction the noncongruent type chooses some action $d \in D_R$.\footnote{Given pure retention strategy, generically the noncongruent type does not randomize.} Then he prefers retention to replacement at this action; since his policy payoff is $-x^2$ and retention contributes $R > 0$, he chooses the lowest action in $D_R$ to minimize policy loss subject to securing retention. That is, the noncongruent type plays $\min D_R$ with probability one. Since $|W| \geq 2$, the congruent type plays $\min D_R$ with probability strictly less than one. Thus $\min D_R$ is chosen more frequently by the noncongruent type than by the congruent type, so the principal's posterior at $\min D_R$ is strictly below $\pi$. By sequential rationality, the principal replaces at $\min D_R$, contradicting $\min D_R \in D_R$.
 
Hence the noncongruent type is always replaced. Being replaced regardless of action and minimizing policy loss $-x^2$, he chooses $\min D$.
\end{proof}
 
\subsection{Results for Section \ref{Section:baseline}}
 
\begin{proof}[Proof of Proposition \ref{Prop~UD}]
We proceed in two steps:

\textbf{1: An uninformative equilibrium exists and is unique.}
 
\textit{Existence.} Conjecture that both types choose $x = 1$ without acquiring information, and the principal retains after observing $x \geq 1$. We specify beliefs and verify incentives.
 
 On path, the principal's posterior at $x = 1$ is $\pi$, justifying retention. Off path at $x = 0$: the divinity criterion considers which types most benefit from a deviation to $x = 0$. The two types most likely to gain are $(c, 0)$ and $n$: each prefers $x = 0$ to $x = 1$ if the retention probability exceeds $1 - 1/R$. Divinity therefore assigns $P(t = c | x = 0) = \frac{\pi p}{\pi p + (1 - \pi)} < \pi$, so the principal replaces. Off path at $x = r$: only type $(c, r)$ benefits from this deviation. Divinity assigns $P(t = c | x = r) = 1$, so the principal retains.
 
 Given these beliefs, neither type profitably deviates without acquiring information. The noncongruent type prefers $x = 1$ (yielding $R - 1$) to $x = 0$ (yielding $0$, since replaced) whenever $R \geq 1$, which holds by Assumption \ref{Assumption: Office}; $x = r$ is strictly dominated for him. The uninformed congruent type prefers $x = 1$ to $x = 0$ and $x=r$ because this is the ex ante optimal action with retention benefits. So no type deviates within the uninformed strategy.
 
Next, the congruent type cannot profitably deviate by acquiring information. If informed, his optimal decision is $x= \omega$ for $\omega\in \{1,r\}$ and $x=1$ if $\omega=0$.  The net benefit from being informed is $p(r-1)^2 = k^0$. For $k > k^0$, this benefit does not exceed the information cost, so the congruent type does not deviate.

\textit{Uniqueness.} We first rule out other pooling profiles. If $x = 1$ leads to replacement, the noncongruent type deviates to $x = 0$, so pooling at $x = 1$ with replacement cannot be part of an equilibrium. Since $x = r$ is dominated for the noncongruent type, the only other pooling candidate is $x = 0$. But pooling at $x = 0$ cannot be an equilibrium either. If $x = 0$ leads to replacement, then $x = 1$ is off path; by divinity its posterior is one, since type $(c,1)$ is most likely to benefit from deviating to $x = 1$, so the principal retains there and the congruent type deviates. If instead $x = 0$ leads to retention, the same deviation to $x = 1$ is again retained and strictly preferred by the congruent type. Either way, pooling at $x = 0$ fails. 

We next rule out separating profiles. Any such profile 
would reveal types, so the noncongruent type is replaced along the path. Since such a  type must choose within $\{0,1\}$, the only plausible action is that the noncongruent type chooses $x=0$. In such an equilibrium, the principal must replace at $x=1$ to prevent the noncongruent type from deviating. However, this is impossible: either $x=1$ is chosen by the congruent type on path, or it is off path, but type $(c,1)$ most benefits from such a deviation, and divinity criterion assigns posterior 1. In both cases, the principal should retain after observing $x=1$, contradiction.

Hence the pooling equilibrium at $x = 1$ is unique among uninformative equilibria.

\textbf{Step 2: No informative equilibrium exists.}
 
Suppose for contradiction an informative equilibrium exists for $k > k^0$. By Lemma \ref{L2}, the noncongruent type chooses $\min D = 0$ and is replaced. We show this leads to a contradiction.

We first argue that all three actions are on path. By Lemma \ref{L2}, the noncongruent type chooses $x = 0$ and is replaced, so $x = 0$ is on path. The informed congruent type chooses $x=\omega$ for $\omega\in \{1,r\}$ regardless of retention strategy. So all $x\in D$ is chosen on path. 
 
By sequential rationality and the posteriors, the principal retains if and only if $x\geq 1$. Given this retention rule, the informed congruent type chooses $x = 1$ when $\omega = 0$, along with $x=\omega$ for $\omega\in \{1,r\}$. The expected gain from being informed is $p(r-1)^2 = k^0$. For $k > k^0$, the congruent type strictly prefers to deviate to staying uninformed and choosing $x = 1$, contradicting the informative equilibrium assumption.
\end{proof}

\begin{Remark}\label{Re~cost}
Under full delegation, informative equilibria do not exist in pure retention strategies for any $k>0$. Allowing mixed retention changes this conclusion: informative equilibria exist if and only if
$k\le \min\{k^0+p(R-1),k_2\}$, where $k_2=pr^2+(1-2p)(1+(r-1)^2)-(1-p)R$. Thus mixed retention can  expand the range of information acquisition under full delegation. The full characterization of these mixed-retention equilibria is given in Section~3 of the Online Appendix.
\end{Remark}

\subsection{Results for Section \ref{Section:constrain}}
\begin{proof}[Proof of Proposition \ref{prop:sq}]
The proof proceeds in two parts.

 \textbf{1. An informative equilibrium exists if and only
if $k \leq k^S$.}

Applying Lemmas~1--2, in any informative equilibrium the
noncongruent type does not acquire information, chooses
$\min D = 1$, and is replaced. Informativeness places $x = r$ on
path, where the principal's posterior $P(t=c|x=r)=1$. Hence the principal
replaces after observing $x = 1$ and retains after observing
$x = r$, and an informative equilibrium can only take the form
described in the proposition. 

Conditional on this retention rule, a
noncongruent type will choose $x = 1$ and give up retention, because
retention begets an additional policy loss of $r^2 - 1$ that exceeds
the office rent $R$. An informed congruent type will choose $x = r$
if $\omega = r$. When $\omega = 1$, his payoff is $R - (r-1)^2$ from
choosing $x = r$ and $0$ from choosing $x = 1$. Since
$R < (r-1)^2$, the congruent type's optimal decision is $x = 1$ when
$\omega = 1$. When $\omega = 0$, his optimal decision is again
$x = 1$. So, our next focus will be on whether the congruent type
may profitably deviate by staying uninformed.
 
An uninformed congruent type's optimal policy can be either $x = 1$ or  $x = r$. By choosing
 $x = 1$, the congruent type receives a payoff of
$-p - p(r-1)^2$, as it leads to removal. By choosing $x = r$, the congruent type receives a payoff of
$-pr^2 - (1-2p)(r-1)^2 + R$, as it leads to retention. The congruent
type's equilibrium payoff is $-p + pR - k$. Thus, he does not have
incentives to deviate if the equilibrium payoff exceeds that of the
best deviation, which amounts to $k \leq k^S$. Since Lemmas~1--2
leave no other candidate for an informative equilibrium, no
informative equilibrium exists when $k > k^S=\min \{p(R+(r-1)^2), p[r^2-1-R]+(1-2p)[(r-1)^2-R]\}.$ Lemma \ref{lemma: cutoff} establishes
that $k^S > k^0$ for a nondegenerate set of parameters.
 
\textbf{2. uninformative equilibria.}

 Generically, when the principal uses pure retention strategies, 
the agent's best responses are pure strategies. By Lemma~1, the noncongruent type never acquires information. Since the noncongruent type finds $x=r$ strictly dominated regardless of the retention strategy, two profiles are candidates: either both types pool on $x = 1$, or the noncongruent type chooses $x=1$ while the congruent type chooses $x=r$.

\textit{Pooling on $x = 1$.} On path the posterior at $x=1$ is $\pi$, so both retention and replacement are sequentially rational. We'll take them in turn. Off path, the noncongruent type never gains by deviating
to $x = r$, while the type $(c,r)$
gains for any retention probability since $0 > R - (r-1)^2$. The divinity criterion thus
assigns $P(t = c|x = r) = 1$. 

Now we examine the agent's equilibrium condition. 
\begin{itemize}
    \item If $x=1$ leads to retention, neither type deviates within uninformed play: doing so reduces policy payoff without retention benefit (both types prefer $x=1$ to $x=r$ when uninformed).  Finally, consider the congruent type's deviation to acquiring
information. If informed, he chooses $x = 1$ for
$\omega \in \{0, 1\}$ and $x = r$ for $\omega = r$, so the value of
information is $p(r-1)^2 = k^0$. The deviation is unprofitable if
and only if $k > k^0$. This gives a pooling equilibrium at $x=1$ with retention, which exists if and only if $k > k^0$.
\item If $x=1$ leads to replacement, 
 the uninformed congruent type prefers $x=1$ replaced to $x=r$ retained if and only if $R - pr^2 - (1-2p)(r-1)^2 \geq -p - p(r-1)^2$, which amounts to
$R \leq \underline{R}:=p(r^2-1)+(1-3p)(r-1)^2$. In this case, the value of acquiring information is $p(R+(r-1)^2)$. For this deviation to be unprofitable requires $k>p(R+(r-1)^2)$. Finally, note that under $R \leq \underline{R}$, $k^S$ simplifies to $k^S = p(R+(r-1)^2)$.
This gives an uninformed pooling equilibrium: it exists if and only if $k > k^S$ and $R \leq \underline{R}$.
\end{itemize}

 \textit{Separation.} Separation reveals types along the
path: the principal replaces after observing $x = 1$ and retains
after observing $x = r$. The noncongruent type does not deviate to
$x = r$. The uninformed congruent type prefers $x = r$
with retention to $x = 1$ with replacement if and only if $R \geq \underline{R}$. His remaining deviation is to acquire
information. Given the retention rule, an informed congruent type
chooses $x = \omega$ when $\omega \in \{1,r\}$;  when $\omega = 0$, his optimal decision is
again $x = 1$. His expected payoff from being informed is
$-p + pR$, so the value of information relative to staying
uninformed at $x = r$ is
$(-p + pR) - \left(R - pr^2 - (1-2p)(r-1)^2\right) =
p[(r^2 - 1) - R] + (1-2p)[(r-1)^2 - R]$, the second component of
$k^S$. Note that under $R \geq \underline{R}$, 
$k^S $ simplifies to $ p[(r^2 - 1) - R] + (1-2p)[(r-1)^2 - R]$. Therefore, whenever
$R \geq \underline{R}$, the deviation to acquiring information is
unprofitable if and only if $k > k^S$. This gives an uninformed separating equilibrium: it exists if and only if $k > k^S$ and
$R \geq \underline{R}$.
 
To summarize, for $k \in (k^0, k^S]$ the informative
equilibrium coexists only with the pooling equilibrium (a). For $k > k^S$, the informative equilibrium is gone and, besides (a), exactly one further uninformed equilibrium survives: the separating equilibrium (b) if $R \geq \underline{R}$, or the replacement-pooling equilibrium (c) if $R \leq \underline{R}$.
\end{proof}

\begin{proof}[Proof of Lemma \ref{lemma: cutoff}] 
 $k^S\geq k^0\Leftrightarrow p[(r^{2}-1)-R]+(1-2p)[(r-1)^{2}-R]\geq p(r-1)^2$, which simplifies to $(1-2p)[(r-1)^2-R]\geq p(2-2r+R)$. This inequality is most likely to hold if $R$ is arbitrarily close to $1$. In this case, it simplifies to $(1-2p)r(r-2)> p(3-2r)$, which is true 
 for any $r\geq 2$.  Thus by continuity,  for $R$ larger than but close to $1$, the inequality holds. This means that a nondegenerate set of $(p,r,R)$ ensures the second component of $k^S$ exceeds $k^0$. The first component satisfies $p(R+(r-1)^2) > p(r-1)^2 = k^0$ trivially, so $k^S = \min\{\cdot,\cdot\} > k^0$ on this set.
\end{proof}

\begin{proof}[Proof of Proposition \ref{prop:m}]
We proceed in two steps.

\textbf{1. An informative equilibrium exists if and only if $k\leq k^M$.}
 
An informed congruent type chooses $x = \omega$ when $\omega \in \{0, r\}$, since the policy stake $r^2$ exceeds the office rent $R$. At $\omega = 1$, the informed congruent type compares $x = 0$ (yielding $-1$ if replaced) with $x = r$ (yielding $R - (r-1)^2$ if retained); he chooses $x = r$ if $R \geq r(r-2)$ and $x = 0$ otherwise. The noncongruent type chooses $x = 0$ in every case (he does not pander to $x = r$ since $R < r^2$). The principal's beliefs are $P(t = c | x = r) = 1$ and $P(t = c | x = 0) < \pi$, so she retains after $x = r$ and replaces after $x = 0$. Given this retention rule, the congruent type's incentive to acquire information is determined by his payoff comparison between informed play and the best uninformed deviation.
 
The uninformed congruent type's optimal action depends on $R$. By choosing $x = 0$ uninformed, he receives $-(1 - 2p) - p r^2$. By choosing $x = r$ uninformed, he receives $-(1 - 2p)(r-1)^2 - p r^2 + R$. The latter is preferred when $R \geq (1 - 2p) r (r - 2)$. The threshold $k^M$ thus takes one of three forms, depending on $R$.
\begin{enumerate}
    \item  If $R \geq r(r - 2)$, the informed congruent type plays $x = r$ for $\omega \in \{1, r\}$ and $x = 0$ for $\omega = 0$, while uninformed play is $x = r$; informed and uninformed payoffs differ only at $\omega = 0$, where the gain from being informed is $r^2 - R$, so $k^M = p(r^2 - R)$.
    \item If $r(r - 2) > R \geq (1 - 2p) r (r - 2)$, the informed congruent type plays $x = r$ for $\omega = r$ and $x = 0$ for $\omega \in \{0, 1\}$, while uninformed play is $x = r$; informed and uninformed payoffs differ at $\omega \in \{0, 1\}$, where the gain from being informed is $r^2 - R$ at $\omega = 0$ and $r(r - 2) - R$ at $\omega = 1$, so $k^M = p(r^2 - R) + (1 - 2p)(r(r-2) - R)$.
    \item  If $R < (1 - 2p) r (r - 2)$, the informed congruent type plays $x = r$ for $\omega = r$ and $x = 0$ for $\omega \in \{0, 1\}$, while uninformed play is $x = 0$; informed and uninformed payoffs differ at $\omega = r$, where the gain from being informed is $r^2 + R$, so $k^M = p(r^2 + R)$.
\end{enumerate}
In each case, the noncongruent type does not deviate to $x = r$: such a deviation would yield $-r^2 + R$, which is strictly negative under Assumption \ref{Assumption: Office}, while his equilibrium payoff at $x = 0$ with replacement is $0$. Conversely, no informative equilibrium exists for $k > k^M$: the structure above is the only candidate (Lemma~\ref{L1}), so the value of information is exactly $k^M$ and information acquisition fails.

\textbf{2. Uninformative equilibria.}
 
We now construct all equilibria in which neither type acquires information. The noncongruent type must choose $x = 0$ since $x=r$ is dominated. The uninformed congruent type chooses either $x = r$ or $x = 0$. This leaves two profiles: separation, with the congruent type at $x = r$ and the noncongruent type at $x = 0$; and pooling, with both types at $x = 0$.
 
\emph{Separation.} The noncongruent type at $x = 0$ reveals himself and is replaced; the congruent type at $x = r$ has posterior one and is retained. The noncongruent type does not deviate to $x = r$, and the uninformed congruent type prefers $x = r$ with retention to $x = 0$ with replacement if and only if $R \geq (1-2p)r(r-2)$. His remaining deviation is to acquire information, which is valued at $p(r^2 - R)$ if $R \geq r(r-2)$ and $p(r^2 - R) + (1-2p)(r(r-2) - R)$ if $r(r-2) > R \geq (1-2p)r(r-2)$; in either case this equals $k^M$. Hence this equilibrium exists if and only if $R \geq (1-2p)r(r-2)$ and $k > k^M$.
 
\emph{Pooling at $x = 0$.} Both types choose $x = 0$, so its posterior is $\pi$ and the principal is indifferent; both retention and replacement are sequentially rational on path. Off path, divinity assigns $P(t=c|x=r)=1$, so the principal retains at $x = r$.
\begin{itemize}
    \item If the principal retains at $x = 0$, the uninformed congruent type prefers $x = 0$ to $x = r$ since $-(1-2p) \geq -(1-2p)(r-1)^2$, and the noncongruent type prefers $x = 0$. Both comparisons are independent of $R$, and an informed congruent type would play $x = r$ only at $\omega = r$, so his value of information is $p r^2$. Hence this equilibrium exists if and only if $k > p r^2$.
    \item If the principal replaces at $x = 0$, the uninformed congruent type prefers $x = 0$ with replacement to $x = r$ with retention if and only if $R \leq (1-2p)r(r-2)$, and the noncongruent type prefers $x = 0$. An informed congruent type would play $x = r$ at $\omega = r$ and be retained, so his value of information is $p(r^2 + R)$, which equals $k^M$ in this range. Hence this equilibrium exists if and only if $R \leq (1-2p)r(r-2)$ and $k >k^M$.
\end{itemize}
 
 Collecting these cases: the separating and replacement-pooling equilibria arise only above the informative-equilibrium cutoff $k^M$ in their respective $R$-regions, so neither coexists with the informative equilibrium. The only uninformed equilibrium that coexists with it over a nontrivial interval is the pooling equilibrium at $x = 0$ with retention, whose cutoff is $p r^2$. Coexistence occurs exactly when $p r^2 < k^M$, that is, when
\[
R < \frac{(1-2p)r(r-2)}{1-p};
\]
for larger $R$ the retention-pooling cutoff lies above $k^M$, and there is no coexistence.
 \end{proof}

Notably, the critical value of information after the elimination of the moderate policy is at least $p(r^2-R)$, which can be higher or lower than the critical value under full delegation, $p(r-1)^2$. This result suggests that the \citet{szalay2005economics} solution, eliminating the moderate option, does not necessarily increase a careerist congruent type's value of information.

\begin{proof}[Proof of Lemma \ref{lemma: SvM}]
Under $R \geq r(r-2)$, Proposition \ref{prop:m} gives $k^M = p(r^2 - R)$. From Proposition \ref{prop:sq}, $k^S = \min\{A, B\}$ with
\[
A := p(R + (r-1)^2), \qquad B := p[(r^2-1) - R] + (1-2p)[(r-1)^2 - R].
\]
Direct calculation yields $A - k^M = p(2R - 2r + 1)$ and $B - k^M = (1-2p)[(r-1)^2 - R] - p$. Thus $A > k^M$ if and only if $2R > 2r-1$, and $B > k^M$ if and only if $p < (1-2p)[(r-1)^2 - R]$. Since $k^S = \min\{A, B\}$, we have $k^S > k^M$ if and only if both conditions hold. The first holds when $R > r - 1/2$; the second holds for $p$ sufficiently small given Assumption \ref{Assumption: Office}.
\end{proof}
\begin{proof}[Proof of Proposition \ref{prop:ext}]

\textbf{1: An uninformed pooling equilibrium exists and is unique.}
 
Conjecture that both types choose $x = 1$ without acquiring information, and the principal retains after $x = 1$ and replaces after $x = 0$. On path, the posterior at $x = 1$ is $\pi$, so retention is sequentially rational. Off path at $x = 0$, divinity assigns $P(t = c | x = 0) = \pi p / [\pi p + (1-\pi)] < \pi$, so the principal replaces. Given this retention rule, neither type deviates without acquiring information. The noncongruent type prefers $x = 1$ with retention to $x = 0$ with replacement since $R \geq 1$. The uninformed congruent type prefers $x = 1$ with retention to $x = 0$ with replacement, since $x = 1$ minimizes ex ante quadratic loss and also delivers $R$. 
The congruent type does not deviate by acquiring information. If informed, he chooses $x = 1$ at every state. Information has zero value, and any $k > 0$ deters acquisition.

Finally, the pooling equilibrium is unique among uninformative equilibria. If $x = 1$ leads to replacement, the noncongruent type deviates to $x = 0$, so pooling at $x = 1$ with replacement is not an equilibrium. Pooling at $x = 0$ fails as well: $x = 1$ is off path, and since type $(c,1)$ is most likely to benefit from deviating there, divinity assigns $P(t = c |x = 1) = 1$, so the principal retains at $x = 1$ and the congruent type deviates. Finally, an uninformed separating profile would reveal type, implying that the congruent type is retained while the noncongruent type is replaced on path. However, whichever action $x\in \{0,1\}$ is chosen on path, the noncongruent type can always deviate to the other one to secure retention. So uninformed separating equilibria are impossible. 
 
\textbf{2: No informative equilibrium exists.}
 
Suppose for contradiction an informative equilibrium exists under $D = \{0, 1\}$. By Lemma \ref{L1}, the noncongruent type does not acquire information; by Lemma \ref{L2}, he chooses $\min D = 0$ and is replaced. Since $x = 1$ is then chosen only by the (informed) congruent type, its posterior is one and the principal retains; $x = 0$ is replaced. But under this retention rule the informed congruent type plays $x = 1$ at every state. Information thus has zero value, so no $k > 0$ supports acquisition.
\end{proof}

\subsection{Results for Section \ref{Section:endogenous delegation}}
\begin{proof}[Proof of Corollary \ref{cor:sq vs full}]
We argue that eliminating the status quo weakly dominates full delegation regardless of which equilibrium is selected, and strictly dominates whenever the informative equilibrium of Proposition \ref{prop:sq} is selected.
 
Assume $k \in (k^0, k^S)$. Under full delegation, Proposition \ref{Prop~UD} establishes that no informative equilibrium exists; both types pool at $x = 1$. Under $D = \{1, r\}$, by part (2) of Proposition \ref{prop:sq} the pooling equilibrium in which both types choose $x = 1$ and are retained coexists with the informative equilibrium.
 
If the pooling equilibrium is selected, both types choose $x = 1$, yielding the same payoff as under full delegation. If the informative equilibrium is selected, the congruent type chooses $x = 1$ for $\omega \in \{0, 1\}$ and $x = r$ for $\omega = r$, while the noncongruent type chooses $x = 1$ without acquiring information. The principal's policy payoff strictly increases, and her selection payoff is also higher, $[(1 - \pi p)\pi + \pi p]B > \pi B$. Hence eliminating the status quo weakly dominates full delegation, and strictly so whenever the informative equilibrium is selected.
\end{proof}

\begin{proof}[Proof of Corollary \ref{cor:sq vs m}]
We compute the principal's total payoff under each restriction and
take differences. When $x=1$ is eliminated, 
a noncongruent type always chooses $x=0$. A congruent type chooses $x=\omega$ for $\omega\in \{0,r\}$, $x=r$ if $R\geq r(r-2)$ and $x=0$ if $R<r(r-2)$ when $\omega=1$. The principal can perfectly identify a congruent type after observing $x=r$; otherwise, she replaces and the replacement is congruent with probability $\pi$. Her total payoff is 
$$ -\pi p_1(r-1)^{2}-(1-\pi)(p_1+p_rr^{2})+[(1-\pi (1-p_0))\pi +\pi (1-p_0)]B, \quad R\geq r(r-2)$$
and 
$$-p_1-(1-\pi)p_rr^2+[\pi(1-\pi p_r)+\pi p_r]B, \quad R< r(r-2).$$

 When $x=0$ is eliminated, a noncongruent type always chooses $x=1$. A congruent type chooses $x=1$ for $\omega\in \{0,1\}$ and $x=r$ for $\omega=r$. The principal can perfectly identify a congruent type after observing $x=r$; otherwise, she replaces and the replacement is congruent with probability $\pi$. Thus, her total payoff is $$-p_0-(1-\pi)p_r(r-1)^{2}+[(1-\pi p_r)\pi +\pi p_r]B.$$

\emph{Case $R < r(r-2)$.} The selection payoffs coincide. Thus, the comparison rests on the policy payoff difference  $$\Pi(p_0,p_1,p_r,r,\pi)=(p_1-p_0)+(1-\pi)p_r(2r-1).$$
which is nonnegative if and only if $p_0\leq \bar{p}_0^{\mathrm{m}}(p_1,p_r,r,\pi)=p_1+(1-\pi)p_r(2r-1)$.  

\emph{Case $R \geq r(r-2)$.} The policy difference is
\[
\Pi(p_0,p_1,p_r,r,\pi) = -p_0 + \pi p_1 (r-1)^2 + (1-\pi)\bigl(p_1 + p_r(2r-1)\bigr).
\]
The selection difference  is
\[
\bigl[\pi(1-p_0) + (1-\pi(1-p_0))\pi\bigr]B
  - \bigl[\pi p_r + (1-\pi p_r)\pi\bigr]B
  = \pi(1-\pi)(1 - p_0 - p_r)B
  = \pi(1-\pi)p_1 B,
\]
where the last equality uses $p_0 + p_1 + p_r = 1$: eliminating $x=1$ separates the congruent type at $\omega = 1$ in
addition to $\omega = r$, a selection advantage worth
$\pi(1-\pi)p_1 B \geq 0$. The total welfare difference in favor of $x=0$ is $\Pi - \pi(1-\pi)p_1 B$, which is nonnegative
if and only if
\[
p_0 \leq \bar{p}_0^{\mathrm{m}}
  := \pi p_1 (r-1)^2 + (1-\pi)\bigl(p_1 + p_r(2r-1)\bigr)
  - \pi(1-\pi)p_1 B.
\]
The threshold is
positive if and only if
\[
B \leq \bar{B}
  := \frac{\pi p_1(r-1)^2 + (1-\pi)\bigl(p_1 + p_r(2r-1)\bigr)}
          {\pi(1-\pi)p_1};
\]
for $B > \bar{B}$, eliminating $x=1$ is preferred.
\end{proof}
 
 \begin{proof}[Proof of Proposition \ref{prop:nonaccountability}]
 Throughout we compare the informative equilibrium under status-quo elimination with delegation to an unaccountable agent. After delegating to an unaccountable official, with probability $\pi$ the principal encounters a congruent type who acquires information and always chooses $x = \omega$; with probability $1 - \pi$ she encounters a noncongruent type who always chooses $x = 0$ without acquiring information. The principal's policy payoff is thus $-(1-\pi)[p_r r^2 + p_1]$, and since she must find a replacement, her selection payoff is $\pi B$.

When the status quo is eliminated, the principal obtains total payoff
\[
-p_0 - (1-\pi)p_r(r-1)^2 + \bigl[(1-\pi p_r)\pi + \pi p_r\bigr]B.
\]
Taking the difference, eliminating the status quo is preferred if and only if
\[
p_0 - (1-\pi)\bigl(p_r(2r-1) + p_1\bigr) \leq \pi(1-\pi)p_r B,
\]
that is, if and only if
\[
p_0 \leq (1-\pi)\bigl(p_r(2r-1) + p_1\bigr) + \pi(1-\pi)p_r B. \qedhere
\]
\end{proof}
 
\section{Robustness}

\subsection{Asymmetric distribution of prior}\label{Asymmetry}
In this subsection, we continue to suppose that the moderate option is the principal's ex ante preferred policy, but drop the  symmetric prior distribution assumption. Our goal is to examine to what extent the symmetric distribution assumption drives the agent's policymaking behaviors. We work with a weaker assumption: 
\begin{Assumption}\label{A1: CP}
The moderate option, $x=1$, is the principal's ex ante preferred policy: 
$2r-1\geq \frac{p_0-p_1}{1-p_0-p_1}$ and $\frac{r+1}{r-1}\geq \frac{1-p_0-2p_1}{p_0}$. 
\end{Assumption}

Below we will present Propositions \ref{Prop~UD}-\ref{prop:ext} with the general distribution of prior. We omit proofs if the new propositions are the same with the benchmark up to a relabeling of parameter values. We will not focus on comparing cutoffs. In fact, since Lemma \ref{lemma: cutoff} shows that $k^S> k^0$ for a nondegenerate set of parameters, and Proposition \ref{prop:m} shows that $k^M> k^0$  for a nondegenerate set of parameters, a continuity argument suggests that a distribution sufficiently close to being ``symmetric'' (i.e., $(p_0\approx p, p_1\approx 1-2p)$) 
supports the same ranking of cutoffs with the baseline.

\begin{propositionp}{\textbf{1}$^*$}\label{prop:gUD}
Under full delegation, i.e., $D=\{0,1,r\}$, an informative equilibrium does not exist whenever $k > p_r(r-1)^2$. A unique equilibrium exists in which both types choose $x=1$ and are retained on path.
\end{propositionp}
Propositions~\ref{prop:gUD} and~\ref{Prop~UD} coincide except that $\omega=r$ occurs with probability $p_r$ rather than $p$, so the value of information changes to $p_r(r-1)^2$; in particular the uniqueness argument of Proposition~\ref{Prop~UD} applies. Proofs are omitted.

\begin{propositionp}{\textbf{2}$^*$}\label{prop:ge}  Let $\hat{k}^S:=\min\{p_r(R+(r-1)^{2}),\, p_0[(r^{2}-1)-R]+p_1[(r-1)^{2}-R]\}$ and $\underline{R}^{*} := p_0(r^2-1) + (p_1 - p_r)(r-1)^2$. When the principal eliminates the status quo, i.e., $D=\{1,r\}$:
(1) an informative equilibrium exists in the subgame if and only if $k \leq \hat{k}^S$. In this equilibrium, the noncongruent type chooses $x=1$ without acquiring information, and the congruent type acquires information and chooses $x=1$ for $\omega\in\{0,1\}$ and $x=r$ for $\omega=r$. Moreover, $\hat{k}^S > k^0$ for a nondegenerate set of parameters.
(2) uninformative equilibria can arise. (a) A pooling equilibrium exists if and only if $k > p_r(r-1)^2$, where both types choose $x=1$ and are retained. (b) A separating equilibrium exists if and only if $k > \hat{k}^S$ and $R \geq \underline{R}^{*}$, where the noncongruent type chooses $x=1$ and is replaced while the congruent type chooses $x=r$ and is retained. (c) A pooling equilibrium exists if and only if $k > \hat{k}^S$ and $R \leq \underline{R}^{*}$, where both types choose $x=1$ and are replaced.
 \end{propositionp}

\begin{proof} Part (1) and the equilibrium constructions in part (2) follow the proof of Proposition~\ref{prop:sq} under the relabeling $p \mapsto p_0$ at $\omega=0$, $1-2p \mapsto p_1$ at $\omega=1$, and $p \mapsto p_r$ at $\omega=r$. We only verify the cutoff $\hat{k}^S$. An uninformed congruent type choosing $x=1$ receives $-p_0 - p_r(r-1)^2$ (replaced), and choosing $x=r$ receives $-p_0 r^2 - p_1(r-1)^2 + R$ (retained); his equilibrium payoff is $-p_0 + p_r R - k$. Comparing the equilibrium payoff to the best deviation gives $k \leq \hat{k}^S$. The separating and replacement-pooling equilibria are the two possible scenarios relative to the uninformed congruent type's indifference at $R = \underline{R}^{*}$, exactly as in Proposition~\ref{prop:sq}.
\end{proof}

\begin{propositionp}{\textbf{3}$^*$}\label{prop:gm}
When the principal eliminates the moderate option, i.e., $D=\{0, r\}$, there exists $\hat{k}^M>0$ such that an informative equilibrium exists if and only if $k \leq \hat{k}^M$. In this equilibrium, the noncongruent type chooses $x=0$ without acquiring information, and the congruent type acquires information and chooses $x=\omega$ for $\omega\in\{0,r\}$; at $\omega=1$ he chooses $x=r$ if $R \geq r(r-2)$ and $x=0$ otherwise. Uninformative equilibria also arise: the congruent type chooses $x=r$ and is retained if $R \geq p_1 r(r-2) + (p_0-p_r)r^2$; otherwise both types pool at $x=0$, where the principal may either retain or replace on path.
\end{propositionp}

\begin{proof}
The proof of Proposition~\ref{prop:m} applies under the relabeling $p \mapsto p_0$ at $\omega=0$, $1-2p \mapsto p_1$ at $\omega=1$, and $p \mapsto p_r$ at $\omega=r$: the informed congruent type's play, the retention rule, the noncongruent type's equilibrium condition, and the  value of information for all parameter values are all unchanged. The only difference is the uninformed congruent type's optimal policy: choosing $x=0$ yields $-p_1 - p_r r^2$ and choosing $x=r$ yields $-p_1(r-1)^2 - p_0 r^2 + R$, so he prefers $x=r$ if and only if $R \geq p_1 r(r-2) + (p_0-p_r)r^2$. This threshold replaces $(1-2p)r(r-2)$ and shifts the boundary between the second and third regions of $R$, giving
\[
\hat{k}^{M} =
\begin{cases}
p_0(r^2 - R), & r(r-2) \leq R < (r-1)^2, \\
p_0(r^2 - R) + p_1(r(r-2) - R), & p_1 r(r-2) + (p_0 - p_r) r^2 \leq R < r(r-2), \\
p_r(r^2 + R), & 1 \leq R < p_1 r(r-2) + (p_0 - p_r) r^2.
\end{cases}
\]
The construction of all uninformative equilibria for $k > \hat{k}^M$ follows the proof of Proposition~\ref{prop:m} under the same relabeling.
\end{proof}

 Proposition  \ref{prop:gm} and \ref{prop:m} are qualitatively similar. 
 
\begin{propositionp}{\textbf{4}$^*$}
Suppose $k > p_r(r-1)^2$. When the principal eliminates the radical reform, i.e., $D=\{0,1\}$, a pooling equilibrium exists in which both types choose $x=1$ without acquiring information and are retained on path.
\end{propositionp}
 
\begin{proof}
Both steps of the proof of Proposition~\ref{prop:ext} apply here with the state probabilities relabeled ($p \mapsto p_0$ at $\omega=0$ and $p \mapsto p_r$ at $\omega=r$): Lemmas~\ref{L1} and~\ref{L2} hold for any prior, the case analysis over retention rules is unchanged, and under the only surviving rule (retain at $x=1$, replace at $x=0$) informed and uninformed play coincide at every state, so information has zero value and any $k>0$ deters acquisition.
\end{proof}

Eliminating the radical reform (i.e., $D=\{0,1\}$) cannot motivate information acquisition when full delegation cannot. This delegation situation is uninteresting.

\textbf{Summary.} Dropping the symmetric prior distribution assumption does not affect the policymaking behavior of an informed congruent or noncongruent type. It only slightly alters the congruent type's value of information, as $p_0$ and $p_r$ may differ. Since the agent's behavior remains qualitatively unchanged, the principal-optimal delegation strategy is also unaffected, aside from relabeling the probabilities of each state. We omit the details.

\subsection{Modeling noncongruence}\label{Noncongruence}
In the main model, the ``noncongruent type'' is so politically conservative that he avoids any policy changes, and that information about the state of the world is worthless to him. Technically, our main results continue to hold if the noncongruent type (1) is more conservative than the congruent type, and (2) places a lower value on information than the congruent type. This specification allows for the existence of a separating equilibrium where the congruent type acquires information and acts informatively, while the noncongruent type remains uninformed and chooses lowest action within the delegation set. Below we consider two modifications.

\subsubsection{Noncongruent type that is overly ``reformist''}
We model a noncongruent type as being more radical than the principal regarding reform decisions. Specifically, his policy payoff is $v_n(x,\omega)=-(x-r)^2$, implying his ideal policy is $x=r$ regardless of the state. As in the baseline model, information is not valuable to this type. 
 
We remark that this modeling choice, i.e., noncongruence as a diehard reformist, seems less reasonable than as a diehard conservative. It is easier to interpret a captured type or ideologue as someone who resists change from the status quo, rather than as someone who resists change from radical reforms.
 
Crucially, we emphasize that this modeling choice is equivalent to the baseline up to a relabeling of variables. We can set $x'=r-x$ and $\omega'=r-\omega$ and work with $(x',\omega')$ instead of $(x,\omega)$. The only difference from the baseline is that the compromising choice shifts from $x=1$ to $x'=r-1$. This will not affect any substantive results if we modify the assumption to $R\in ((r-1)^2, 1)$ with $r<2$.
 
\subsubsection{Noncongruent type that values information}
We assume a noncongruent type  with the following state-dependent preferences regarding policymaking: his policy payoff is $v_n(x,\omega)=-(x-x_n^*(\omega))^2$, where the noncongruent type's bliss policy in state $\omega$ is
\begin{align*}
    x_n^*(\omega)=\begin{cases}
        0, & \omega\in \{0,1\}\\
        1, & \omega=r
    \end{cases}.
\end{align*}
An informed noncongruent type's ideal policy is lower than the congruent type's bliss $x_c^*(\omega)=\omega$. Moreover, an uninformed noncongruent type's preferred policy is $x=0$, lower than an uninformed congruent type's $x=1$. Although information is valuable to the noncongruent type, his value of information, $p$, is strictly lower than that of the congruent type, $p((r-1)^2+1)$.

Our analysis proceeds with assuming symmetric distribution of priors ($p_0=p_r=p<1/2$) and relatively high cost of information ($k>p(r-1)^2$). 
To simplify the algebra, we slightly modify the assumption on office rent:
\begin{assumptionp}{\textbf{2}$^n$}
        $R\in (\max \{1, r(r-2)\}, (r-1)^2)$
\end{assumptionp}
We will  replicate Proposition \ref{Prop~UD}-\ref{prop:ext}, focusing on characterizing PBEs in each delegation scenario where the principal uses pure retention strategies. This restriction implies that generically the agent's best responses are also pure strategies.   
 
We call an event ``good'' (bad) news for retention if the principal's posterior belief about congruence conditional on this event increases (decreases) relative to prior. Throughout, the divinity type space includes information states, as in Appendix A: a hypothetical informed type's equilibrium payoff is evaluated at the action his equilibrium strategy prescribes for the corresponding uninformed type, and the sunk information cost does not enter the comparison.

The following lemma replaces Lemma 1 of the main paper, which does not apply here since the noncongruent type's preferences are state dependent. 

\begin{lemma}\label{lem:n-no-info}
In any equilibrium with pure retention under any $D\subseteq\{0,1,r\}$, the noncongruent type does not acquire information if $k>p$.
\end{lemma}
\begin{proof}
Singleton delegation sets are trivial.

Suppose first $1\in D$. We argue that the noncongruent type never chooses $x=r$, informed or not. To see this, 
choosing $x=r$ implies a policy loss of at least $(r-1)^2$ relative to choosing $x=1$, which offsets the retention benefit $R$. This means that the noncongruent type's essential choice set is $D\cap\{0,1\}$. If this set is singleton, the value of information is $0$, so we suppose  both actions $\{0,1\}$ are available. The noncongruent type would match the state when $|\rho(1)-\rho(0)|=0$. In this case, the value of information is exactly $p$. The noncongruent type would pander to the action leading to retention if $|\rho(1)-\rho(0)|=1$. In this case, the value of information is $0$. 

Suppose instead $D=\{0,r\}$. We argue that information may matter to the noncongruent type only at $\omega=r$. To see this, by staying uninformed, the noncongruent's payoff is $\rho(0)R-p$ from choosing $x=0$ and $\rho(r)R-[(1-p)r^2+p(r-1)^2]$ from choosing $x=r$. Choosing $x=0$ over $x=r$ brings a payoff gain of $r(r-2p)-(\rho(r)-\rho(0))R\geq r(r-1)-R>0$. This means that the uninformed noncongruent type's optimal action is $x=0$. By acquiring information, he would choose $x=0$ for $\omega\in \{0,1\}$ and $x=r$ for $\omega=r$, since the retention benefit $R$ offsets the policy loss $(r-1)^2-1$. Thus, the noncongruent type's value of information is $p(R-r(r-2))$, which is less than $p$. 

In all cases the value of information is at most $p<k$.
\end{proof}

%We remark that the bound is not vacuous under $D=\{0,r\}$: when $R>r(r-2)$, the profile in which both types acquire information and pool their pandering at $x=r$ is belief consistent, with posterior $\tfrac{\pi(1-p)}{\pi(1-p)+(1-\pi)p}>\pi$ at $x=r$; only the acquisition constraint of Lemma \ref{lem:n-no-info} eliminates it. The bound also relies on pure retention: with interior retention probabilities the gap on $\{0,1\}$ can fall below the unit policy stake, and the noncongruent type's value of information can exceed $p$. 

\textbf{Full delegation}
\begin{propositionp}{\textbf{1}$^n$}\label{prop:nud}
Under full delegation, i.e.,  $D=\{0,1,r\}$, an informative equilibrium does not exist whenever $k> p(r-1)^2$. A unique equilibrium exists where  both types of the agent choose  $x=1$ and are retained. 
\end{propositionp}
\begin{proof} We proceed in two steps.

\textbf{1. An uninformative equilibrium exists and is unique.}

We first show that the strategy specified as above can be supported in an equilibrium.  Applying the refinement, we note: [1] The type $(c,r)$ is the most likely deviator to $x=r$, so the divinity criterion assigns $P(t=c|x=r)=1$. [2]  The type $(c,0), (n,0), (n,1)$ are the  most likely deviators to $x=r$, and divinity assigns $P(t=c|x=0)=\frac{\pi p}{\pi p+(1-\pi)(1-p)}<\pi$, so the principal replaces. The principal's sequentially rational retention strategy is to replace after observing $x=0$ and retain after observing $x=1$ and $x=r$. 
 
Now we check if any agent may gain by deviating from the pooling strategy. The congruent type cannot profitably deviate for reasons as in the proof of Proposition \ref{Prop~UD}. Consider the noncongruent type. By choosing the status quo, he receives a  payoff of $-p$. By choosing the equilibrium strategy, his payoff is $-(1-p)+R$. Thus, the deviation is unprofitable. The noncongruent type does not want to deviate to acquiring information either, because he will not choose $x=0$ even when it is the ideal policy. 

Next, we rule out all other equilibria in which neither type acquires information.

\textit{Pooling at $x=1$ with replacement.} The noncongruent type deviates to his ex ante optimal action $x=0$. Contradiction.

\textit{Pooling at $x=0$.} Then $x=1$ is off path. The type $(c,\emptyset)$ gains $1+2p(r-2)$ in policy from deviating to $x=1$ and thus requires a retention probability of $1-\frac{1+2p(r-2)}{R}$; the types $(c,1)$ and $(n,r)$ gain only $1$ and require $1-1/R$, which is strictly larger since $r>2$; all other types gain less. So $(c,\emptyset)$ is the unique most likely deviator, divinity assigns $P(t=c|x=1)=1$, and the principal retains at $x=1$. But then $(c,\emptyset)$ deviates, whether $x=0$ is retained or replaced. Contradiction.

\textit{Pooling at $x=r$.} The noncongruent type deviates to $x=0$: even if replaced, this gains $(1-p)r^2+p(r-1)^2-p-R>0$ under Assumption \textbf{2}$^n$. Contradiction.

\emph{Separation.} The noncongruent type is revealed and replaced on path, so his action in this conjectured equilibrium can only be $x=0$. The congruent type is retained at some $a_c\neq 0$. If $a_c=1$, the noncongruent type can profitably deviate to $x=1$. If $a_c=r$, then $x=1$ is off path; the types $(c,0),(c,1)$ and $(n,r)$ benefit from deviating there at any retention probability.  The divinity assigns $P(t=c|x=1)=\frac{\pi(1-p)}{\pi(1-p)+(1-\pi)p}>\pi$ and the principal retains at $x=1$. The noncongruent type thus can  profitably deviate to $x=1$. Contradiction.

Now we will rule out all informative equilibria. 
 
(1) The noncongruent type does not acquire information,
but the congruent type does.
 
The noncongruent type plays an uninformed strategy, and his expected policy payoffs rank the actions
as in the baseline: $x = 0$ yields $-p$, $x = 1$ yields $-(1-p)$, and
$x = r$ yields $-(1-p)r^2 - p(r-1)^2$, so his ranking is
$0 \succ 1 \succ r$. The proof of Lemma~2 applies here: if the noncongruent type played any action leading to retention, he would play the lowest one with the set. But the informed congruent type chooses it strictly less often, rendering it bad news for retention. Contradiction. The noncongruent type is therefore replaced and chooses
$x = 0$. Given this, Step~2 of the proof of Proposition~1 applies here, as the remaining argument involves only the congruent type's
payoffs which are unchanged: all three actions are on path, sequential
rationality ensures retention after $x = 1$ and $x = r$. Under this
retention rule the congruent type's value of information is $p(r-1)^2 < k$, contradicting information acquisition.

(2) Both types of the agent acquires information. 
 
Case I: Suppose the principal retains  after observing $x=0$. [a] If she does not retain upon observing $x=1$, then both types of the agent will choose $x=0$ for $\omega\in \{0,1\}$. Moreover, when $\omega=r$, the congruent type chooses $x=r$ 
while the noncongruent type chooses $x=0$. However, the noncongruent type can profitably deviate to not acquiring information. Contradiction. [b] If she also retains after $x=1$, then $x=0$ is chosen strictly more often by the noncongruent type. This renders $x=0$ bad news for retention. Contradiction. 
 
Thus, we have shown that the principal cannot retain after observing $x=0$.
 
Case II: Suppose the principal retains after observing $x=1$ and replaces after observing $x=0$. Then, the congruent type will choose $x=1$ if $\omega\in \{0,1\}$. However, the congruent type should not have acquired information: he will use information only when $\omega=r$, which means that his information value cannot exceed $p(r-1)^2$. He can profitably deviate by staying uninformed and choose $x=1$.  
 
Thus, we have shown that the principal cannot retain after observing $x=1$ and replaces after observing $x=0$.
 
Case III: Suppose the principal retains after observing $x=r$ and replaces after observing $x=0$ and $x=1$. Given this retention strategy,  the congruent type will choose $x=\omega$ for all $\omega$ while the noncongruent type will choose $x=0$ for  $\omega\in \{0,1\}$ and $x=1$ for $\omega=r$. Since the noncongruent type is always replaced, he can profitably deviate to not acquiring information and choosing $x=0$, as $k > p(r-1)^2 \geq p$.

Case IV: Suppose the principal always replaces. Given this retention strategy,  the congruent type will choose $x=\omega$ for all $\omega$ while the noncongruent type will choose $x=0$ for  $\omega\in \{0,1\}$ and $x=1$ for $\omega=r$. But the noncongruent type can deviate to not acquiring information and choosing $x=0$.

Thus, ``both types of the agent acquire information'' cannot be part of an equilibrium.

(3) The noncongruent type acquires information but the congruent type does not. 
 
Case I: Suppose the principal retains after observing $x=0$. Then the noncongruent type will choose $x=0$ when $\omega\in \{0,1\}$. He finds $x=r$ dominated, so he must choose either $x=0$ or $x=1$  when $\omega=r$ to utilize information. However, he can profitably deviate to choosing $x=0$ without acquiring information. Thus, the principal cannot retain after observing $x=0$.
 
Case II: Suppose the principal retains observing $x=1$ and replaces after observing $x=0$. Then the noncongruent type will choose $x=1$ for $\omega\in \{0,1\}$. In this case, since information may be useful to him only if $\omega=r$, the noncongruent type can profitably deviate by not acquiring information and choosing $x=1$.

Case III: Suppose the principal  replaces after observing $x\in \{0,1\}$. Given this strategy, the noncongruent type is sure to be replaced as he never chooses $x=r$. In this case, acquiring information is strictly dominated for him.

Thus we have finished the proof.
\end{proof}
 
 \textbf{Constrained delegation}

From now on, we maintain the assumption $k>k^0=p(r-1)^2$.  
 
\begin{propositionp}{\textbf{2}$^n$}\label{prop:nsq} Suppose $k>k^0$. When the principal eliminates the status quo, i.e., $D=\{1,r\}$,
(1) $\exists k^S>0$ such that an informative equilibrium exists in the subgame if and only if $k\leq k^S$. In this equilibrium, the noncongruent type chooses $x=1$ without acquiring information. The congruent type acquires information and chooses $x=1$ for $\omega\in \{0,1\}$, and $x=r$ for $\omega=r$. Moreover, $k^S>k^0$ for a nondegenerate set of parameters. (2) uninformative equilibria can arise. (a) A pooling equilibrium exists if and only if $k > k^0$, where both types choose $x=1$ and are retained. (b) A separating equilibrium exists if and only if $k > k^S$ and $R \geq \underline{R}$, where the noncongruent type chooses $x=1$ and is replaced while the congruent type chooses $x=r$ and is retained. (c) A pooling equilibrium exists if and only if $k > k^S$ and $R \leq \underline{R}$, where both types choose $x=1$ and are replaced. 
\end{propositionp}
\begin{proof}
The congruent type's incentives are exactly those of Proposition~\ref{prop:sq}, and in every equilibrium below the noncongruent type is uninformed. This is because within $D = \{1,r\}$ information cannot improve his choice: at each state his bliss lies weakly below $x=1$, so he chooses $x=1$ whenever it is retained and $x=r$ is dominated. The characterization therefore coincides with that of Proposition~\ref{prop:sq}.
 
For part (1), the retention rule is to replace after $x=1$ and retain after $x=r$; holding fixed the acquisition decisions, the policy choices in the proposition are optimal, and since no parameter values differ from Proposition~\ref{prop:sq}, the same cutoff $k^S$ supports the informative equilibrium.
 
For part (2), the noncongruent type finds information useless regardless of the retention rule, so he stays uninformed; the congruent type then faces exactly the strategic problem of Proposition~\ref{prop:sq}. Its three uninformative equilibria carry over: pooling at $x=1$ with retention (a), which exists for $k > k^0$; separation (b), which exists for $k > k^S$ and $R \geq \underline{R}$; and pooling at $x=1$ with replacement (c), which exists for $k > k^S$ and $R \leq \underline{R}$.
\end{proof}
Proposition \ref{prop:nsq} resembles Proposition \ref{prop:sq}:  the noncongruent type is unlikely to acquire information when the congruent type does not.

\begin{propositionp}{\textbf{3}$^n$}\label{prop:nm} 
Suppose $k>k^0$. When the principal eliminates the moderate option, i.e., $D=\{0, r\}$, $\exists k^M>0$ such that an informative equilibrium exists in the subgame if and only if $k\leq k^M$. In this equilibrium, the noncongruent type chooses $x=0$ without acquiring information, and the congruent type acquires information and chooses $x=\omega$ for $\omega\in \{0,r\}$; at $\omega=1$ he chooses $x=r$ (the maintained assumption $R > r(r-2)$ places us in this case). Uninformative equilibria also arise: a separating equilibrium in which the congruent type chooses $x=r$ (retained) and the noncongruent type chooses $x=0$ (replaced), for $k > k^M$; and a pooling equilibrium in which both types choose $x=0$ and are retained, for $k > p r^2$.
\end{propositionp}
\begin{proof} 
When $k>k^0$, in any equilibrium the noncongruent type does not acquire information. This is because such a type's value of information is at most $p$, which is below $k^0=p(r-1)^2$. 

\textit{1. The informative equilibrium exists if and only if $k\leq k^M$ and is unique.}

An informative equilibrium must feature the congruent type acquiring information while the noncongruent type does not. The informed congruent type chooses $x=\omega$ for $\omega\in\{0,r\}$, and $x=r$ at $\omega=1$. The uninformed noncongruent type chooses $x=0$ as proved in Lemma \ref{lem:n-no-info}. On path, the posterior at $x=0$ is $\frac{\pi p}{\pi p+(1-\pi)}<\pi$ and the posterior at $x=r$ is $1$, so replacing at $x=0$ and retaining at $x=r$ is sequentially rational. The congruent type's value of information is $p(r^2-R)=:k^M$, so he acquires if and only if $k\leq k^M$. This establishes existence for $k\leq k^M$ and nonexistence of this profile for $k>k^M$.

\textit{2. Uninformative equilibria.} The noncongruent type chooses $x=0$ as proved in Lemma \ref{lem:n-no-info}. The uninformed congruent type chooses either $x=r$ or $x=0$.
 
\textit{Separation.} The noncongruent type is replaced at $x=0$ and the congruent type is retained at $x=r$. Under $R\geq r(r-2)$ the uninformed congruent type prefers $x=r$ retained to $x=0$ replaced, and does not deviate to acquiring information, worth $p(r^2-R)=k^M$. Hence this equilibrium exists for $k > k^M$.
 
\emph{Pooling at $x=0$ with retention.} Both types choose $x=0$, so its posterior is $\pi$ and retention is sequentially rational; off path, divinity assigns $P(t=c|x=r)=1$, so the principal retains at $x=r$. Both types earn a higher payoff from choosing $x=0$ than $x=r$, so they will not deviate. An informed congruent type would play $x=r$ only at $\omega=r$, so his value of information is $p r^2$. Hence this equilibrium exists, whenever $k > p r^2$. Finally, since $p r^2 < k^M$ fails under $R \geq r(r-2)$, it does not coexist with the informative equilibrium here.

Pooling at $x=0$ with replacement  cannot be part of an equilibrium. The uninformed congruent type can profitably deviate to $x=r$ since the retention benefit $R$ more than offset the policy loss  $(1-2p)r(r-2)$. 
\end{proof}
 
 \begin{propositionp}{\textbf{4}$^n$}\label{prop:next}
Suppose  $k> p(r-1)^2$.  When the principal eliminates the radical reform, i.e., $D=\{0,1\}$, no informative equilibrium exists. The unique equilibrium pools both types at $x=1$ without information acquisition; the principal retains after $x=1$ and replaces after $x=0$.
\end{propositionp}
\begin{proof}
By Lemma \ref{lem:n-no-info}, the noncongruent type is uninformed in every
equilibrium. 

\textit{1. No informative equilibrium exists.} 

Suppose the congruent type acquires information on path. If the principal retains after $x=0$, the uninformed noncongruent type chooses $x=0$ with probability one while an informed congruent type chooses it less often, so $x=0$ is bad news for retention. Contradiction. If she replaces after both actions, the noncongruent type chooses $x=0$ while the informed congruent type chooses $x=1$ at $\omega\in\{1,r\}$, so the posterior at $x=1$ is one, contradicting replacement. If she retains after $x=1$ and replaces after $x=0$, the congruent type chooses $x=1$ for all $\omega$, so he does not acquire information. Contradiction.

\textit{2. Uninformative equilibria} 

\textit{Existence.} Conjecture that both types choose $x=1$ without acquiring information, the principal retains after $x=1$ and replaces after $x=0$. On path, the posterior at $x=1$ is $\pi$, justifying retention. Off path at $x=0$, the types $(c,0)$, $(n,0)$, and $(n,1)$ are the most likely deviators. Divinity assigns $P(t=c|x=0)=\frac{\pi p}{\pi p+(1-\pi)(1-p)}<\pi$, justifying replacement. Given this rule, neither type deviates in policy, informed or not. Information therefore has zero value and neither type acquires it.

\textit{Uniqueness.} We exhaust all remaining possibilities. (1) Pooling at $x=1$ with replacement fails, because the noncongruent type deviates to his preferred policy $x=0$. (2) Pooling at $x=0$ fails. $x=1$ is off path, the type $(c,\emptyset)$ gains $1+2p(r-2)$ in policy from deviating there, strictly more than the gain of $1$ to $(c,1)$ and $(n,r)$ since $r>2$, so divinity assigns $P(t=c|x=1)=1$. The principal retains at $x=1$, and $(c,\emptyset)$ deviates there. (3) Separation fails. The noncongruent type is revealed on path, so the only candidate for separation is the noncongruent type choosing $x=0$ with replacement while the congruent type choosing $x=1$ with retention. But the noncongruent type deviates to $x=1$.
\end{proof}

Comparing Propositions \ref{prop:nud}-\ref{prop:next} to Proposition \ref{Prop~UD}-\ref{prop:ext}, we can see that they are substantively similar.

\end{appendices}

\end{document}